**Identifying robust communities and multi-community nodes by combining top-down and bottom-up approaches to clustering**

Chris Gaiteri[1,2*], Mingming Chen[3*], Boleslaw Szymanski[3,4], Konstantin Kuzmin[3], Jierui Xie[3*,5], Changkyu Lee[2], Timothy Blanche[2], Elias Chaibub Neto[6], Su-Chun Huang[7], Thomas Grabowski[7,8], Tara Madhyastha[8] and Vitalina Komashko[9]

1 Rush University Medical Center, Alzheimer's Disease Center, Chicago, IL
2 Allen Institute for Brain Science, Modeling, Analysis and Theory Group, Seattle, WA
3 Rennselaer Polytechnic Institute, Department of Computer Science, Troy, NY
4 Społeczna Akademia Nauk, Łódź, Poland
5 Samsung Research America, San Jose, CA
6 Sage Bionetworks, Seattle, WA
7 University of Washington, Department of Neurology, Seattle, WA
8 University of Washington, Department of Radiology, Seattle, WA
9 Trialomics, Seattle WA
*These authors contributed equally to this work

**Abstract**
Biological functions are carried out by groups of interacting molecules, cells or tissues, known as communities. Membership in these communities may overlap when biological components are involved in multiple functions. However, traditional clustering methods detect non-overlapping communities. These detected communities may also be unstable and difficult to replicate, because traditional methods are sensitive to noise and parameter settings. These aspects of traditional clustering methods limit our ability to detect biological communities, and therefore our ability to understand biological functions.

To address these limitations and detect robust overlapping biological communities, we propose an unorthodox clustering method called SpeakEasy which identifies communities using top-down and bottom-up approaches simultaneously. Specifically, nodes join communities based on their local connections, as well as global information about the network structure. This method can quantify the stability of each community, automatically identify the number of communities, and quickly cluster networks with hundreds of thousands of nodes.

SpeakEasy shows top performance on synthetic clustering benchmarks and accurately identifies meaningful biological communities in a range of datasets, including: gene microarrays, protein interactions, sorted cell populations, electrophysiology and fMRI brain imaging.

**Introduction**
Molecules, cells and tissues carry out biological processes through physical interaction networks [1,2,3] and can enter disease states when those networks are disrupted [4,5,6,7]. Because the structure of networks is related to the functions they carry out [8,9], it is possible to investigate biological functions by examining network structure [3,10,11,12,13,14]. Densely connected groups known as communities are prevalent in biological networks and may be related to specific molecular, cellular or tissue functions [10,15,16,17]. Therefore, biological community detection is a key first step in many network-based biological investigations. However, accurately identifying biological communities is challenging, because network structures often have incorrect or missing links, because traditional methods can produce unstable results [18,19], and because biological communities tend to be highly overlapping [20,21,22].

*SpeakEasy: A new label propagation algorithm to detect overlapping clusters*
We propose a label propagation clustering algorithm, "SpeakEasy", to robustly detect both overlapping and non-overlapping (disjoint) clusters in biological networks. SpeakEasy is related to earlier label propagation algorithms [23,24,25] in the sense that nodes join communities based on exchange of "labels" between connected nodes. These "labels" do not refer to *a priori* community titles. In this context, labels are unique bits of information that are assigned randomly and used to track cluster membership. SpeakEasy differs from previous label propagation algorithms, because nodes update their labels on the basis of their neighbors' labels, while subtracting the expected frequency of these labels, based on their popularity in the complete network. This process combines a bottom-up approach to clustering (using neighboring information) with a top-down approach (using information from the whole network). This dual approach facilitates accurate community detection in many types of biological networks (Table 1) because top-down information is used to ensure the bottom-up label propagation process identifies communities that accurately represent the global network structure [19,26,27,28].

In addition to accurate cluster detection (see Results section), community detection via SpeakEasy has several practical advantages for biological applications. For instance, since the number of communities in a dataset is rarely known in advance, SpeakEasy automatically predicts the number of communities and does *not* require manual tuning of clustering parameters for good results. Second, it can cluster networks with any type of links (weighted/unweighted, directed/undirected, positive/negative-valued edges) or any type of network structure (scale-free or any other distribution of connectivity). SpeakEasy is highly scalable and can cluster networks with hundreds of thousands of nodes. Third, because it is very efficient, the stochastic clustering process can be repeated many times to detect robust clusters that are not muddled by data artifacts or noise. The repeated clustering process also allows SpeakEasy to identify multi-community nodes, whose membership tends to oscillate between different clusters. Finally, users can select overlapping or non-overlapping output, as is appropriate for their applications.

*Visual example of SpeakEasy clustering*

For an intuitive example of how SpeakEasy identifies communities, we illustrate the clustering process on a demonstration network (Figure 1A). This network can represent any type of biological component, such as genes, proteins or tissues; network links could be derived from primary data or scientific literature. Initially, labels (represented by colored tags) are applied randomly to all nodes (Figure 1A). Then, each node updates its label, based on the labels of neighboring nodes. Specifically, a node will adopt the label found most commonly on its neighbors taking into account the global frequency of all labels (i.e., it will adopt the label that is most specific to its neighbors). For instance, the node shown in gray (Figure 1B) is connected to orange-, blue- or green-labeled communities, so it must adopt one of these three labels. The gray node will update its label to the blue tag, because it has the strongest specific connection to the blue community, even though it has an equal number of links to the green community. Through this updating process, densely connected groups of nodes will acquire the same label. Multi-community nodes tend to oscillate their membership between multiple communities, such as the node located between the red and orange communities (Figure 1B).

**Results**
*Summary*
We use three approaches to determine the accuracy of SpeakEasy community detection. First, we test its performance on a large set of synthetic networks with carefully controlled characteristics, wherein the true clusters are known. Then we apply it to real-world networks, wherein the true clusters are unknown (Table 2). In this second context we can quantify community detection accuracy by using the statistical separation between clusters. Finally, we apply SpeakEasy to several types of common biological networks (Table 1). This collection of applications was selected because they have multiple of the following characteristics: 1) analysis of these datasets often utilizes clustering; 2) they have high levels of noise; 3) they are generated via different technologies measuring biological properties at several physical scales; 4) they can benefit from overlapping community detection, and 5) their true community structure is unknown or debated. In all cases, we make comparisons to alternate methods that have been applied to the same or similar datasets.

*Synthetic clustering benchmarks*
To generate networks with known community structure, we use the Lancichinetti-Fortunato-Radicchi (LFR) benchmarks, which are widely used to test overlapping and non-overlapping clustering methods [29]. These benchmarks contains a range of networks, some with well-separated clusters and other networks with clusters that are highly cross-linked and almost indistinguishable. We track the accuracy of communities detected by SpeakEasy under increasing levels of cross-linking ($\mu$) (Figure 2A). The effect of cross-linking (increasing $\mu$) is reflected by decreasing modularity (Q) and modularity density ($Q_{ds}$) (Figure 2B). SpeakEasy shows the highest yet accuracy in community detection, based on normalized mutual information (NMI) [25,30,31,32,33], especially for highly cross-linked clusters ($\mu=0.95$) (Figure 2A). Additional cluster recovery statistics such as the adjusted Rand index have varying inputs and sensitivity [34], but also support this strong ability to detect true communities. These results are not affected by various distributions of cluster size or intra-cluster degree distributions (Figure S1). Thus, SpeakEasy can accurately identify disjoint clusters in the most popular clustering benchmarks, even when these clusters are heavily obscured by cross-linking/noise.

We also test community detection on LFR networks with overlapping communities. In this setting, SpeakEasy also shows excellent community detection performance and the ability to identify multi-community nodes (Figure 2C, 2D) [35]. As seen previously for disjoint networks (Figure 2A), increasing the level of cluster cross-linking (μ) makes community detection more challenging, resulting in lower NMI with the true set of clusters. Better community detection accuracy was achieved for networks with higher average connectivity (D). This can be explained by the greater cluster density of these networks (Figure 2). Community detection is also affected by the number of communities that are tied to multi-community nodes ($O_m$). When multi-community nodes are tied to many communities (high $O_m$ values), community detection becomes more difficult (Figure 2C, 2D). This response to highly overlapping communities is universal across overlapping clustering algorithms [35]. Community detection scores for most methods also tend to decrease on large networks [35]. This decrease in performance could be more severe for SpeakEasy, because it employs a diffusion process. However, SpeakEasy performs slightly better on networks of 5000 nodes versus networks with 1000 nodes. This may be explain by the incorporation of global network information (label popularity) into the local clustering process [26,27,28].

*Abstract clustering performance on diverse real-world networks*
The LFR benchmarks accurately represent certain aspects of social and biological networks, but are limited in other aspects. For example, networks in the LFR benchmarks have low transitivity and null assortativity (propensity for hubs to connect to hubs) [36]. Therefore we apply SpeakEasy to fifteen real networks that are often used to test clustering methods. Unlike the LFR benchmarks, the true community memberships in these networks are unknown. However, the quality of clusters detected by various methods can be compared by using modularity (Q) [37] and modularity density scores ($Q_{ds}$) [38], which quantify how well a given network is segmented into dense clusters.

We compare modularity values from SpeakEasy to those another label propagation algorithm, GANXiS, because that method showed the best overlapping clustering performance in a recent comparison of clustering methods [35]. In this comparison, SpeakEasy shows improved performance on 6 out of 15 networks using the modularity (Q) metric, with a mean percent difference in performance of 2% over GANXiS (Table 2). Using the more accurate $Q_{ds}$ metric that corrects two well-known flaws in the original Q metric [38,39], SpeakEasy performs better than GANXiS on 14 out of 15 networks with a mean percent difference of 28% over GANXiS (see Supplementary Materials). The consistently high $Q_{ds}$ values from SpeakEasy (compared to Q-values) indicate that it tends to detect more small and highly dense clusters than GANXiS [38]. SpeakEasy shows both higher Q and $Q_{ds}$ scores for the two biological networks in this test set ('dolphins' and 'c.elegans'). These modularity values are approach those of methods that directly attempt to maximize modularity [34]. Consistently high modularity on networks of diverse origin indicates that a simultaneous top-down and bottom-up approach to clustering functions will on a wide range of topologies. However, high modularity is still not a proof of real utility in clustering biological networks. Therefore, we apply SpeakEasy to several types of biological networks, and compare the output clusters to gold-standards or to literature-based ontologies.

*Application to protein-protein interaction datasets*
Because a single protein may be part of more than one protein complex (set of bound proteins that work as a unit), Discovery of protein complexes directly benefits from development of methods which detect overlapping communities. We test SpeakEasy community detection of overlapping protein complexes, using two well-studied high-throughput protein interaction networks (Gavin et al. [40] and Collins et al. [41]) derived from affinity purification and mass spectrometry (AP-MS) techniques. We then compare the predicted clusters against three gold-standards for protein complexes [42,43,44] (Figure 3). NMI scores between the predicted and the true protein complexes indicate that SpeakEasy produces the most accurate recovery of protein complexes to date [32,33,45] (Table 3). We also examine precision and recall statistics specifically for the detection of multi-community nodes. SpeakEasy identifies a smaller number of multi-community nodes than are listed in various gold-standards, although the multi-community nodes it does detect are often in agreement with the gold-standards (Table 3). However, there may be upper limits on using the Collins and Gavin datasets to measure multi-community node detection, because there is frequently no evidence (links) in these networks in support of canonical multi-community nodes (Figure 3 inset).

*Application to cell-type clustering*
Identifying robust cell populations that constitute a true cell type is a challenging problem, due to ever-increasing levels of detail on cellular diversity. To explore how traditional clustering methods and SpeakEasy can be used to identify robust cell-types, we use a collection of sorted cell populations from the Immunologic Genome Project (Immgen) [46,47]. The immune system contains many populations of cells that can be distinguished by specific combinations of cell surface markers as well as broader functional families, such as dendritic cells, macrophages and natural killer cells. We apply SpeakEasy to a matrix of expression similarity from cells from 212 cell types, as defined in Immgen. We then compare our results with the primary classification of the sorted cells. There is a strong correspondence between the identified clusters and the tissue origin of these cells. (Figure 4, Table 4).

We find that applying SpeakEasy once again, to each of these broad categories of cell types, identifies sub-communities with higher correspondence to the tissue of origin and cell type, considered together (Table 4). Thus, successive applications of SpeakEasy clustering results may reflect successive tiers of biological organization. In comparison to standard hierarchical clustering methods, even when those methods are supplied with the true number of clusters, SpeakEasy still shows the highest correspondences with canonical cell types (see Supplementary Materials). These results indicate

SpeakEasy will be useful in future applications, where the number of communities (in this case, cell types) is unknown.

*Application to finding coexpressed gene sets*
Several cellular or molecular processes can generate correlated gene expression (called coexpression), including cell-type variation, transcription factors, epigenetic or chromosome configuration [48]. Identifying genes which are coexpressed in microarray or RNAseq datasets is useful because these gene sets may carry out some collective functions related to disease or other phenotypes. This task is challenging because coexpressed genes may be context-specific and therefore lack gold-standards, gene expression data tends to be noisy, and these gene sets are generated by overlapping mechanisms [21,49].

Therefore, we use SpeakEasy to detect overlapping and non-overlapping coexpressed gene sets in two datasets that are commonly used to address many biological questions: The Human Brain Atlas (HBA) [50], comprised of 3584 microarrays measured in 232 brain regions and the Cancer Cell Line Encyclopedia (CCLE) [51], comprised of 1037 microarrays from tumors found in all major organs. We find 40 non-overlapping clusters in HBA containing more than 30 genes (a practical threshold to assess functional enrichment), with a median membership of 384 (see Supplementary Materials). In CCLE we find 43 clusters with more than 30 gene members, with a median community size of 265. Coexpressed gene sets tend to be involved in certain biological functions; therefore, these gene sets tend to have high functional enrichment scores based on ontology databases such as Gene Ontology (GO) and Biocarta [50]. Of these 40 large clusters we detect in HBA, 27 have an average Bonferroni-adjusted p-value of <0.01 for one or more biological processes. Of the 43 large clusters we detect in CCLE, 35 have a Bonferronni-adjusted p-value of <0.01.

We also generate overlapping clusters from both the HBA and CCLE datasets. Overlapping coexpressed gene sets may be useful in biological studies because gene coexpression is driven by overlapping mechanisms [21]. Furthermore, assigning truly multi-community nodes to only a single community will produce inherently inaccurate communities. When multi-community SpeakEasy output is enabled, we still detect 40 clusters in HBA data, but the median size increases from 384 to 544, with 4510 genes holding overlapping community membership. Overlapping results from CCLE show an increase in median module size from 265 (non-overlapping) to 702, with ~10,000 genes found in more than one community. Functional enrichment scores for overlapping HBA gene sets are equivalent to non-overlapping results, while enrichment scores for gene sets from CCLE were several orders of magnitude more significant. We conduct a comparison of these results to the WGCNA method commonly used to identify coexpressed genes (see Supplementary Material), which shows practical benefits of SpeakEasy, including higher functional enrichment and avoiding of arbitrary filters and complex parameter settings.

*Application to neuronal spike sorting*
Extracellular neuronal recording with single electrodes, tetrodes, or high density multichannel electrode arrays can detect the activity of multiple nearby neurons. However, these combined responses must be separated into responses of specific neurons. This blind source separation process is known as "spike sorting", because each spike is assigned to a particular theorized neuron. Single neurons often generate relatively unique signatures (i.e. spike waveform shapes and amplitude distributions on multiple adjacent electrodes), and emerge as clusters in the matrix of waveform correlations.

To realistically test spike sorting, it is important to match noise levels in real brain recordings. Therefore, we use real depth-electrode recordings generate a simulated time-series of spikes in which the true spike times and unique neuronal sources are known (see Supplementary Materials). Comparison of the inferred clusters (represent the activity of a single neuron) to the true associations between spikes and neurons indicates that SpeakEasy can reliably sort spikes from multielectrode recordings (Table S1). The waveforms associated with each cluster can then be used in template-matching to detect additional spikes from the same neuronal origin.

*Application to resting-state fMRI data*
Functional magnetic resonance imaging (fMRI), obtained while a subject is at rest (rs-fMRI), is a valuable tool in understanding of systems-level changes in a variety of domains, including neurodegenerative disease [52]. Correlations between the rs-fMRI signals in different regions of interest (ROIs) may indicate which regions are functionally related. Brain networks composed of functionally-related ROI's tend to be noisy and overlapping because ROIs perform functions for multiple networks or because the low temporal resolution of the blood oxygen level-dependent signal causes temporal smearing of brain networks. The ability to robustly identify functional networks (communities), and changes to this structure that occur with disease, is critical to understanding the physiological changes that may be early indicators of disrupted cognitive function.

Figure 5A shows the relatively small inter-regional correlations characteristic of rs-fMRI functional connectivity graphs in control subjects (n=21) and subjects with Parkinson disease (PD, n=27) [53] (Table S2). Due to high levels of noise and weak community structure (Figure 5A), apparent communities of brain regions may easily be driven by clustering parameters or data artifacts. Therefore, we apply SpeakEasy to the average control and PD rs-fMRI connectivity matrices 1000 times, to quantify the stability of each cluster through co-occurrence matrices (Figure 5B). For instance, in control subjects, the community of temporal areas is very stable (has high average co-occurrence) while the cluster of parietal areas is less stable. We then use a permutation test to identify communities of brain regions that change their membership between control and PD groups (see Supplementary Materials).

Communities identified in control and PD groups contain similar sets of brain regions (NMI=.51) (Table S3), but the specific communities do alter their membership significantly in PD.  Using clusters from control subjects as a frame of reference, we observe both significant changes in community size and inter-community connectivity (see Supplementary Materials).  A cluster comprised of (predominantly) temporal cortex ROIs showed the largest drop (-27%) in average co-occurrence among its members in PD ($p<0.001$).  Specifically, the temporal cluster disintegrated in PD, with its area-members joining different communities (Figure 5B).  In PD subjects, the putamen and thalamus regions form an independent cluster in PD that is not observed in the control subjects, wherein those regions are part of the third largest cluster that is composed of temporal and occipital locations regions.  Comparing these results to the alternative clustering method, Infomap [54], which has been used previously with fMRI data [55], show that method is sensitive to arbitrary link thresholds that it requires (see Supplementary Materials and Table S3).

**Discussion**
Biological communities are a common feature of biological networks [9,10] and are associated with execution of various cellular and molecular functions [12,14,15,56].  Therefore, identifying these communities with clustering methods is often the first step in understanding biological datasets.  An ideal clustering algorithm should identify correct clusters in a synthetic setting and have excellent modularity results when true communities are unknown.  Moreover, it should run in a reasonable time on large networks using standard hardware and without the need to manually "tune" method parameters for good results.  When applied to biological networks, it should function well regardless of the type of data or particular network properties of the dataset.  Finally these results should be robust and not driven by noise or method parameters.  SpeakEasy is designed to fulfill these criteria.  Using a wide range of networks (Table 2) SpeakEasy produces higher modularity density scores than the best performing overlapping clustering method to date [25,35].  It has excellent absolute and relative performance on the LFR benchmarks (Figure 2), scales well and can quickly cluster networks with hundreds of thousands of nodes on a typical laptop (Table 1 and Supplementary Materials).  When applied to biological networks generated by distinct experimental methods, SpeakEasy detects robust, plausible, well-validated clusters (Figures 3-5, Tables 3-4, Supplementary Materials).  Collectively these results point to future potential for robust disjoint and overlapping clustering in related applications.

The SpeakEasy algorithm could potentially be improved by changing how node labels are updated. Currently, nodes are updated to reflect the single most unexpected label among their neighbors. However, each node could be simultaneously characterized by multiple unexpected labels.  This might aid in the identification of multi-community nodes or completely nested networks.  In addition, binomial or multinomial tests may provide more accurate metrics for the unexpectedness of a given label. However, this altered label selection would not extend easily to weighted networks or networks with negative link weights.  Selecting an updated label from a randomly chosen subset of inputs could improve results, as analogous improvements have been observed in Bayesian network inference when nodes have greater freedom to reconfigure their local network [57].  With these potential modifications, care must be taken to ensure that the network still converges to a clustered solution and does not become chaotic.

SpeakEasy could also be improved by altering the consensus clustering routine used to identify the final partition and multi-community nodes.  This consensus clustering step is completely separable from the label propagation process.  Therefore, improvements to consensus clustering method could improve the overall results of SpeakEasy.  An ideal consensus clustering method would quickly refine the structure of all of the clusters, using all partitions and output disjoint or overlapping clusters.  However, few available techniques meet these criteria and consensus cluster methods are often slower than primary clustering methods [18,58,59].

While SpeakEasy shows top performance among other available methods on multiple benchmarks and biological datasets, some alternative algorithms produce more accurate results for high $O_m$ values on the LFR benchmarks [35].  However, the exact structure of a network is typically unknown in advance of clustering.  Therefore, the generally excellent performance of SpeakEasy across many simulated and real networks indicates it will likely produce useful results on many datasets in the future.

**Methods**
*Synthetic network benchmarks*
To robustly measure the ability of SpeakEasy to recover true clusters from a range of network structures in the LFR benchmarks, we vary network characteristics (Figure 2, Figure S1) including number of nodes, density of connections, distribution of cluster sizes, cluster separation and number of overlapping communities (see Supplementary Materials).

*Algorithm overview*
An implementation of the SpeakEasy algorithm is provided free for non-commercial use here: doi:… (available on publication) and it is also presented in pseudo-code here (see Supplementary Materials). In summary, initially each node is assigned a random unique label.  Then for some number of iterations (usually less than 30), each node updates its status to the label found among nodes connected to it which has the greatest specificity, i.e. the label with the greatest difference between the actual and the expected frequency (Figure 1 and Supplementary Materials).  Positively or negatively-weighted links between nodes (often produced when clustering correlation-based networks) are easily incorporated into SpeakEasy, as they provide relative increases or decreases in the popularity of a particular label.  The label updating step is performed simultaneously for all nodes.  Although there is the potential for oscillating states to emerge with a simultaneous update step, in practice this is not observed in SpeakEasy.  Cluster accuracy improves when labels from the last several time-steps are included in the calculation of expected and actual labels.  However, initially the network has no history of labels, so we create an artificial buffer of random neighboring labels. This buffer prevents the algorithm from becoming trapped in an early equilibrium, and also provides unique initial conditions, which are useful when clustering the same dataset multiple times.

*Defining disjoint and overlapping communities*

Stochastic clustering algorithms such as SpeakEasy can generate many partitions (sets of clusters) from repeated runs with different initial conditions. Combining these partitions (consensus clustering) is a challenging mathematical process, potentially even more difficult and computationally intensive than clustering individual elements [18,58]. While many consensus clustering techniques attempt to identify the optimal partition, we choose to define a final set of clusters in a way that is representative of the distribution of partitions. Specifically, the partition with the highest average adjusted Rand index (ARI) among all other partitions is selected as the representative partition. Clusters identified in this way are likely to be robust, because spurious partitions will have lower ARI scores with most other partitions. Multi-community nodes are selected as nodes which co-occur with more than one of the final clusters with greater than a user-selected frequency (see Supplementary Materials).

**Funding Acknowledgments:**
This work was supported in part by the Army Research Laboratory under Cooperative Agreement Number W911NF-09-2-0053, the Office of Naval Research Grant No. N00014-09-1-0607 and the National Institutes of Health 1RC4NS073008-01. The funders had no role in study design, data collection and analysis, decision to publish, or preparation of the manuscript. No additional external funding was received for this study.

**Table and Figure Legends**

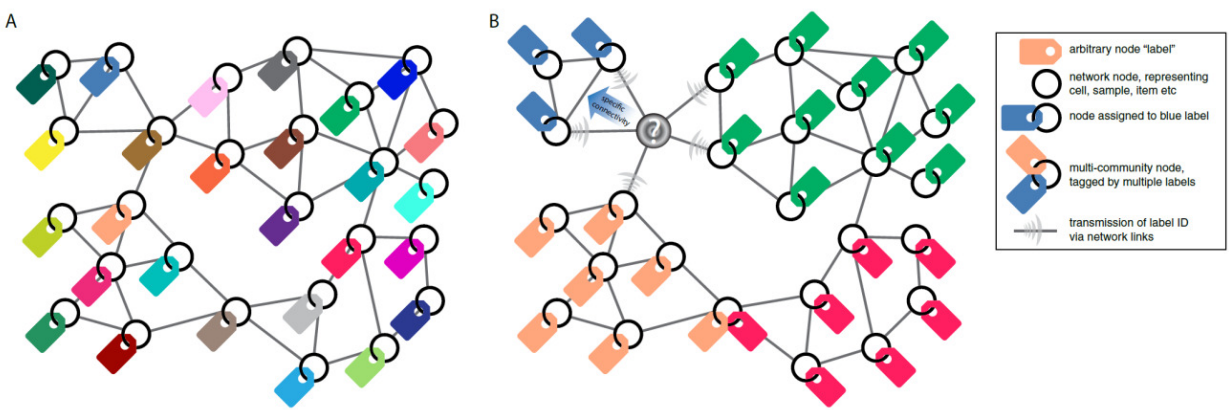

**Figure 1. Intuitive schematic of the core SpeakEasy clustering mechanism.** (A) Clusters are determined by competition between nodes through "labels" (symbolized here by colored tags) that grow and spread through a network. (B) SpeakEasy groups nodes according to the communities to which they are most specifically connected. Thus, when nodes connected to the gray node broadcast their identities, it will join the "blue" community on the upper left, because its connectivity to more popular labels is greater than expected at random. Nodes are classified as multi-community nodes if they fit equally well with multiple communities (for example, node tagged with both orange and red labels).

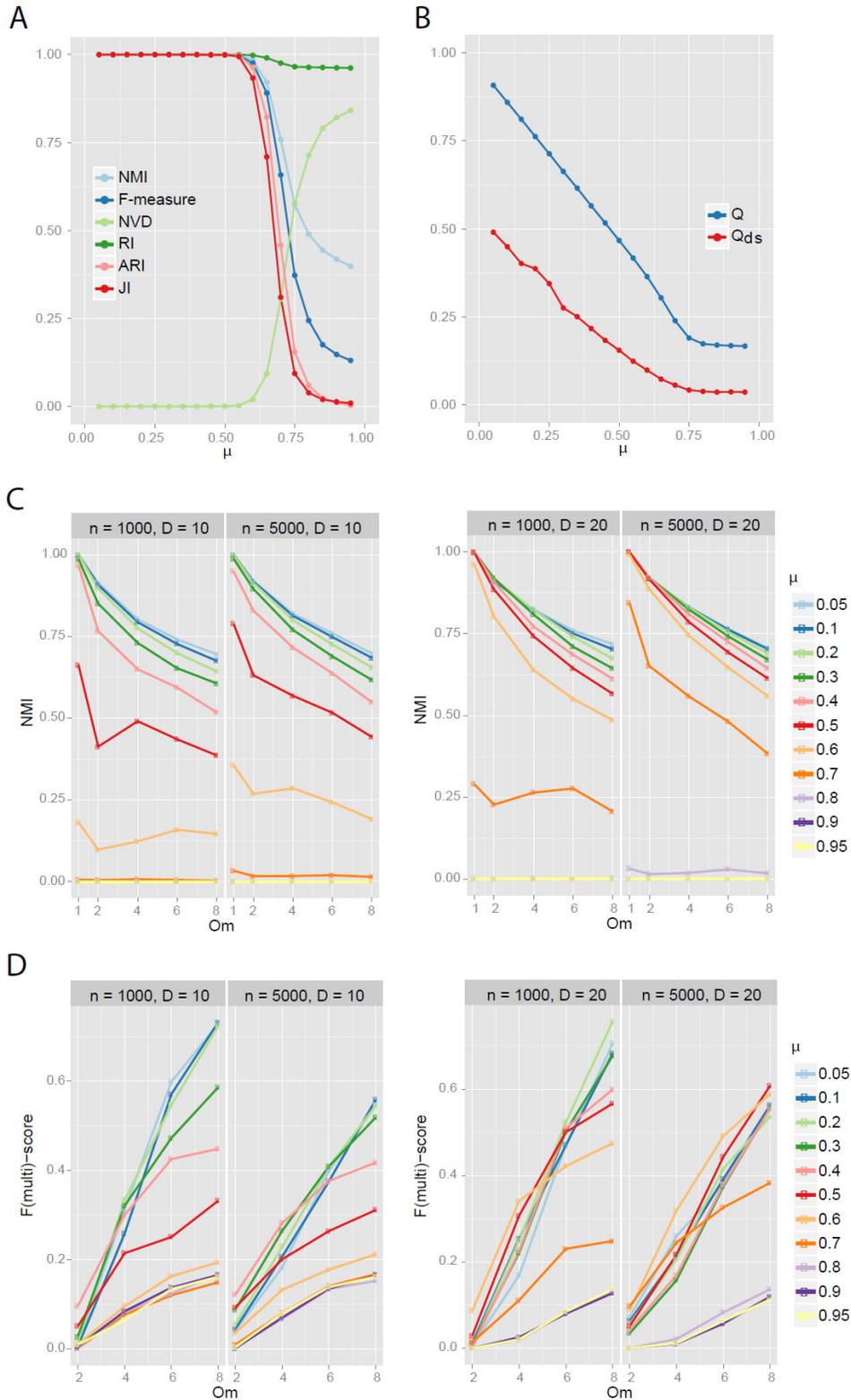

**Figure 2. Disjoint cluster detection performance.** (A) The LFR benchmarks track cluster recovery as networks become increasingly cross-linked (as µ increases) for γ (cluster size distribution parameter) equal to 2 and β (within-cluster degree distribution

parameter) equal to 1. Several metrics characterize cluster recovery with varying levels of sensitivity. For the following measures (min=0), lower values indicate better alignment between the true partition and partition generated by SpeakEasy: NVD - Normalized Van Dongen metric. For the following measures, larger values (max=1) indicate better alignment between the true and SpeakEasy partitions: NMI - Normalized Mutual Information; F-measure; RI- Rand Index; ARI - Adjusted Rand Index; JI - Jaccard Index. See Chen et al. [34] for additional details on these statistical measures. (B) These modularity values provide a statistical estimate of the separation between clusters. For both Q (modularity) and $Q_{ds}$ (modularity density), larger values (max=1) indicate better community separation. (C) Recovery of true clusters quantified by NMI as a function of μ (cross-linking between clusters) and $O_m$ (number of communities associated with each multi-community node). (D) F(multi)-score is the standard F-score, but specifically applied for detection of correct community associations of multi-community nodes, calculated at various values of $O_m$ and different average connectivity levels (D=10,20). NMI metric used for overlapping communities (panels C,D) does not reduce to disjoint NMI, so NMI scores for Om=1, cannot be directly compared to panel A.

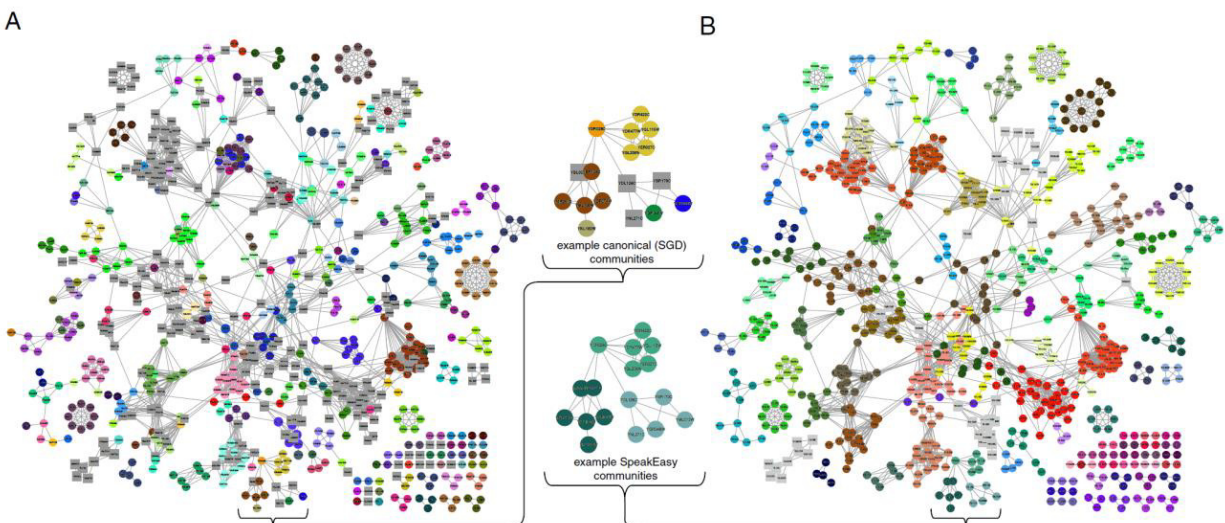

**Figure 3. Contrasting protein complex membership, estimated by small-scale experiments and high-throughput clustering.** (A) The high throughput interaction dataset from Gavin et al. [39] has nodes colored according to complexes found in the Saccharomyces Genome Database (SGD) database. Nodes found in multiple protein complexes are shown as gray squares. (B) The clusters identified by SpeakEasy are color-coded. Nodes found in multiple communities are depicted as gray squares. Inset: network fragments show example positions of actual versus inferred multi-community nodes in a portion of the network, showing how some canonical multi-community nodes have very little support for that classification, based on the network structure.

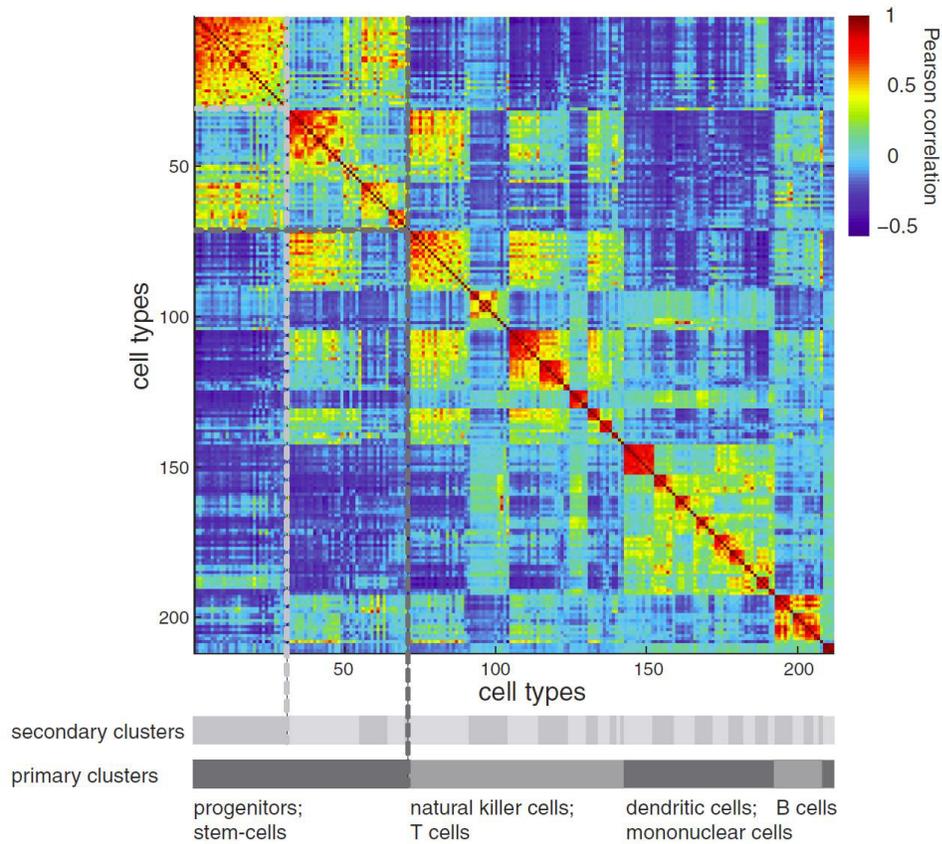

**Figure 4. Primary and secondary biological classifications of immune cell types are reflected in primary and secondary clusters.** The clustered correlation matrix of similarity of cell expression vectors is ordered according to primary clusters, which correspond to large-scale cell families such as B-cells, and secondary clusters, which correspond more closely to a more detailed classification of the intersection of cell-type and tissue of origin (see also Table 4).

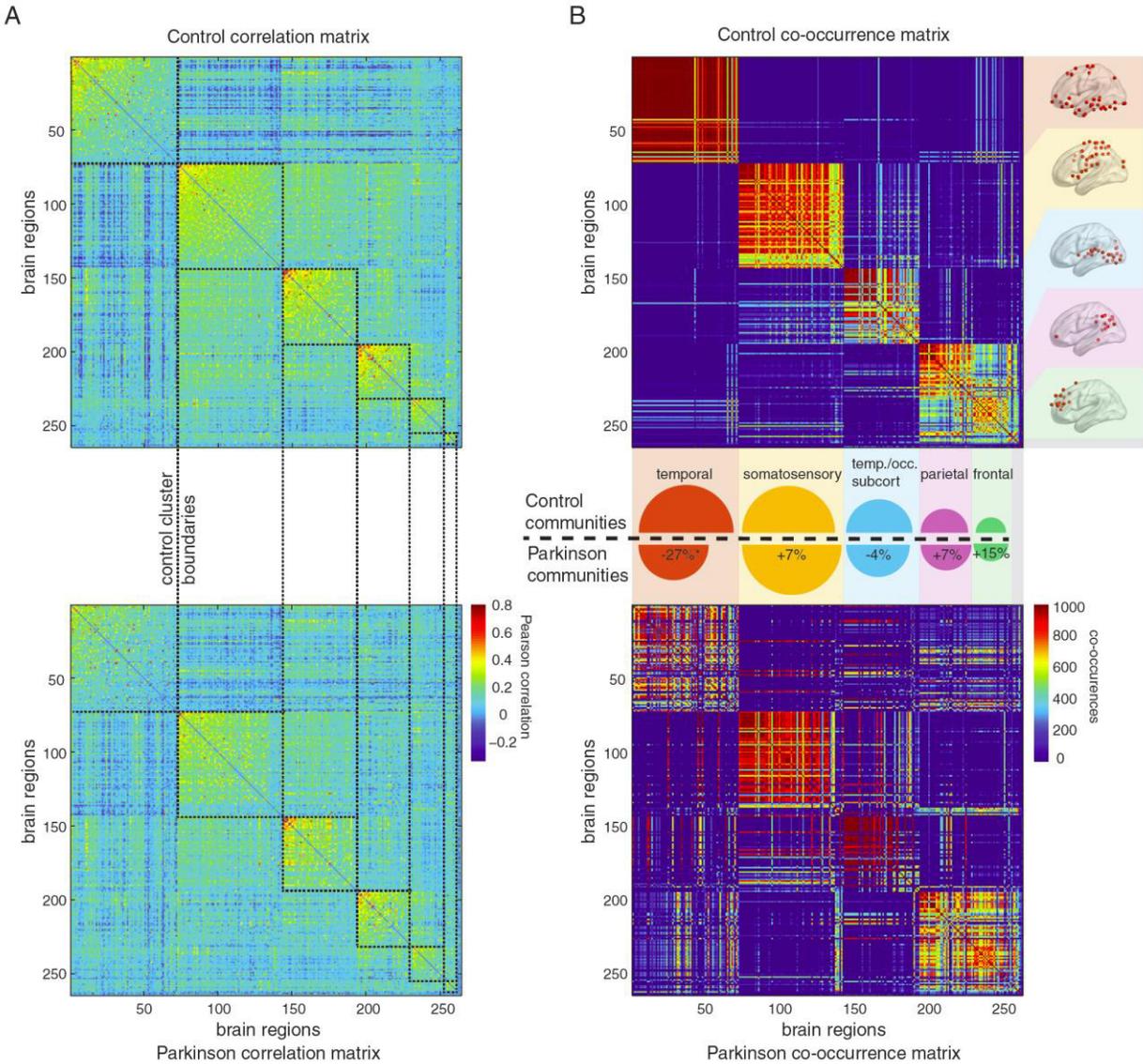

**Figure 5. Shifts within and between resting-state brain communities in Parkinson disease.** (A) Raw correlation matrices between resting state brain activity from control and Parkinson disease cohorts. Dashed lines indicate clusters identified by SpeakEasy from control-state data. Order of brain regions is identical in all matrices (reflects control-state clusters). (B) Co-occurrence matrices for controls and Parkinson disease cohorts. Entries in co-occurrence matrices count the number of times nodes (i,j) are found together in 100 replicated clustering results. (Inset) Semi-circles are scaled by volume to cluster size in control data. The *difference* in size of the corresponding lower semi-circles illustrates the change in average co-occurrence for each control-state cluster. Thus smaller semi-circles in disease (lower half) denote loss of coherence among members of a particular cluster. Text in semi-circles summarizes the most common regional characteristic of each cluster.

**Table 1. Overview of datasets used in SpeakEasy community detection.** We test community detection across a range of biological datasets to robustly characterize the ability to define practically useful biological communities.

**Table 2. Comparison of the abstract goodness of clustering results using modularity (Q and $Q_{ds}$) on many types of networks between SpeakEasy and a top-performing overlapping clustering method (GANXiS).** By testing community detection in many types of networks we can assess the quality of SpeakEasy community detection across networks with different topologies. Top modularity scores are shown in bold. "Karate" is a network of friendships between college club participants from the 1970's. "Pol books" is a co-purchasing network of books on political topics that were published in 2004. "Netscience" is a co-citation network among network science authors. "Dolphins" is a social interaction network of a bottlenose dolphin pod from New Zealand. "Les Miserables" is a network of character interactions in the novel by Victor Hugo. "Football" is a network of American Division 1A college football teams, linked by matches. "Sante Fe" is a co-authorship network of members at the Santa Fe Institute. Links in the "Jazz" network denote musical collaborations between the years 1912 and 1940. "Pol blogs" is a network of hyperlinks among political-oriented blogs in 2005. "Email" is a network of emails linking various Enron employees. The PGP network describes Pretty Good Privacy key signing. "DBLP" is a co-authorship network in computer science, whose communities tend to be related to specific conferences or journals. "Amazon" is a network of item co-purchases.

**Table 3. Comparison between protein complexes defined by small-scale experiments versus those inferred from high-throughput interaction datasets.** Table values consist of normalized mutual information (NMI) between predicted and canonical protein complexes.

**Table 4. Comparison of clusters and subclusters of gene expression vectors from sorted cell populations to canonical families of mouse immune cell-types.** Table values consist of normalized mutual information (NMI) between predicted and canonical protein complexes, for hierarchical clustering methods with various levels of linkage and numbers of clusters.

## Supplemental Methods

**Data sources for synthetic and non-biological networks**

*Synthetic network benchmarks*
For disjoint community detection, we used the LFR benchmarks with 1000 nodes with average degree (number of connections) of 15 and maximum degree of 50 (Figure 2A). The exponent, γ, for the degree sequence varies from 2 to 3 and the exponent, β, for the community size distribution, varies from 1 to 2. Results over four pairs of the exponents (γ, β) = (2, 1), (2, 2), (3, 1), and (3, 2) indicate community detection is robust over this parameter range (Figure 2A, Figure S1). For each of these community distributions, the fraction of between- versus within-community connections (µ) was varied from 0.05 to 0.95 (Figure 2, Figure S1). This means that each node shares a fraction (1- µ) of its edges with the nodes in its community and a fraction µ of its edges with the nodes outside its community. Thus, low µ-values indicate the test networks which are composed of relatively isolated communities, and which should be relatively easy to accurately define. To generate robust results, 10 network instances are generated for each value of µ and we report average performance across a variety of statistics commonly used to assess community recovery (Figure 2A), including NMI, ARI and F-measure [1].

When networks contain multi-community nodes, whose links are evenly divided between multiple communities, community detection is even more challenging. To generate overlapping networks in the LFR benchmarks, we follow the standard practice of setting 10% of the nodes to have their connection evenly divided between two or more communities ($O_m$ parameter) (Figure 2C). This form of overlapping is distinct from the random between-community connections (parameterized by µ), because this 10% subset of nodes is equally well connected to multiple communities. We report average community recovery, using a variety of statistics, averaged over 10 network instances for each value of $O_m$, across a variety of community mixing levels (µ) (Figure 2D). Not all traditional community quality metrics extend to overlapping networks, so we implement an extension of the normalized mutual information (NMI) and the Omega Index (OI) to track recovery of overlapping communities [1]. The OI is equivalent to the common Adjusted Rand Index (ARI) for disjoint communities, while the overlapping NMI measure does not reduce to the standard formulation of NMI for disjoint communities.

These overlapping community detection results could primarily be driven by SpeakEasy's excellent disjoint community performance, since most nodes have a single non-overlapping true community assignment. Therefore, we specifically track recovery of multi-community nodes using a statistic we call F(multi)-score, which is computed identically to the standard F-score, but the inputs specifically track if multi-community nodes are correctly assigned to all of their true communities (Figure 2D). These results indicate that SpeakEasy fulfills the goal of detecting multi-community nodes and does not purely rely on its strong disjoint community detection abilities. Like the related label propagation algorithm, GANXiS [2], SpeakEasy shows a rare upward trend in F(multi)-score as $O_m$ increases. This result, specifically on multi-community nodes (Figure 2D), should be considered with the overall lower community recovery at higher $O_m$ values (Figure 2C). Together, these results indicate that while multi-community nodes increase the difficulty of community detection, it is still possible to cluster such networks accurately and to detect the sets of communities associated with multi-community nodes.

*Abstract clustering performance on real-world networks*
Traditionally, performance of clustering methods on networks with *unknown* correct clustering solutions is measured in terms of modularity ("Q"). Modularity measures the number of within-community connections, relative to the number expected at random [3]. This measure has a maximum value of 1, but in practice maximum possible Q-value will be less than 1.0, due to between-community links. Nevertheless, comparing Q-values between partitions generated by different clustering methods provides a relative measure of their ability to detect well-separated communities. Since the classic modularity formula does incorporate the number of between module connections but not the density of nodes within communities, anomalous situations can occur, wherein better partitions have *lower* Q values (contrary to the intended operation of Q) [4,5]. Therefore, we also track performance of SpeakEasy in terms of an updated modularity measure, $Q_{ds}$, which adds a correction term to the original modularity formula to create a more accurate metric [4].

**Algorithm details**

*Pseudo-code for SpeakEasy*
SpeakEasy(*G*, *numHistoryLabels*)
Parameters:
*G*: The network on which a community structure is to be detected. Input networks may consist of links (values) that are weighted/unweighted, symmetric/directed, positive or positive+negative.
*numHistoryLabels*: Initially, each node is assigned *numHistoryLabels* labels into its historical label buffer.

```
1:    nodes = loadNetwork(G);
2:    numNodes = nodes.size();
3:    // Initialize the historical label buffer of each node as its own node ID.
4:    for i = 0 to numNodes − 1 do
6:        node.historyLabels.add(node[i].nodeId);
7:    end for
8:    // Each node randomly selects (numHistoryLabels − 1) labels from its neighbors
      and add them into its historical label buffer.
9:    for i = 0 to numNodes − 1 do
10:       for j = 0 to numHistoryLabels − 2 do
11:           // A node randomly selects its neighbor and adds the neighbor's initial first
              label into its own historical label buffer.
12:           neighbor = random(node[i].neighbors());
```

```
13:         nbInitialLabel = neighbor.historyLabels[0];
14:         node.historyLabels.add(nbInitialLabel);
15:      end for
16:  end for
17:  // Label propagation procedure of SpeakEasy.
18:  repeat
19:  // Get the global frequencies of all labels in the network.
20:     globalFrequencies.clear();
21:     totalNumLabels = numNodes ∗ numHistoryLabels;
22:     for i = 0 to numNodes − 1 do
23:        for j = 0 to numHistoryLabels − 1 do
24:           label = node[i].historyLabels[j];
25:           globalFrequencies[label]+ = 1/totalNumLabels;
26:        end for
27:     end for
28:  // Update each node's historical label buffer with the most unexpected popular
         label received from its neighbors.
29:     for i = 0 to numNodes − 1 do
30:        node = nodes[i];
31:        // Get the actual numbers of the labels that the node receives from its neighbors.
32:        actualLabelNums.clear();
33:        neighbors = node.neighbors();
34:        while neighbors.hasNext() do
35:           [neighbor weight] = neighbors.next();
36:           for j = 0 to numHistoryLabels − 1 do
37:              label = neighbor.historyLabels[j];
38:              actualLabelNums[label]+= weight;
39:           end for
40:        end while
41:        // Determine the most unexpected popular label.
42:        numNeighbors = neighbors.size();
43:        while actualLabelNums.hasNext() do
44:        // Node computes the actual number of times this label is present at neighbors.
45:           [label actualNum] = actualLabelNums.next();
46:           // Node calculates the expected number of times that this label should be
                present in neighbors based on the global frequency of this label.
47:           expectedNum = globalFrequencies[label] ∗ numNeighbors ∗
                numHistoryLabels;
48:           // Node chooses the label with the largest actual number relative to its
                expected number.
49:           mostUnexpectedLabel =label with max(actualNum − expectedNum);
50:        end while
51:        node.historyLabels.remove(0);
52:        node.historyLabels.add(mostUnexpectedLabel);
53:     end for
54:  until none of the nodes updates its labels for a certain number of iterations
```

*Computational complexity of SpeakEasy*
The initialization takes O(|V| * *numHistoryLabels*) steps to assign an initial buffer with a certain number of history labels to each node.

The label propagation procedure requires O(|V| * *numHistoryLabels*) operations to calculate the global frequencies of all labels in the network.

Getting the actual numbers of labels that a node receives from its neighbors costs O(<k> * *numHistoryLabels*) where <k> is the average degree of the network.

The procedure to determine the most unexpected label for a node has a complexity at most O(<k> * *numHistoryLabels*) because there are at most *numHistoryLabels* distinct labels in the buffer of each of its neighbors.

Thus, it totally requires O(|V| * <k> * *numHistoryLabels*) to update each node's historical label buffer with the most unexpected popular label received from its neighbors. In total, the label propagation procedure has a complexity of O(|V| * <k> * *numHistoryLabels*) which can be expressed as O(|E| * *numHistoryLabels*). Moreover, the label propagation takes a certain number of iteration, denoted as *T*, which is usually a small constant.

Thus, the complexity for the label propagation of SpeakEasy is O(*T* * |E| * *numHistoryLabels*). Since *T* and *numHistoryLabels* are small constants (*T* has a default value of 50 and *numHistoryLables* defaults to 5), the overall complexity can be reduced to O(|E|) which is linear in terms of the network size for sparse matrices.

*Computational complexity of SpeakEasy in practice*
SpeakEasy scales linearly with the number of edges, and can cluster typical biological networks quickly. For instance SpeakEasy clusters 10,000 nodes with random 2% connectivity density (2 million links) in ~10 seconds with an i7 2620M processor, using ~1GB of RAM. Clustering the amazon co-purchase network with over 300,000 nodes (Table 2) required 45 seconds on the same processor and 0.5GB of RAM. The typically sparse connectivity of biological networks [6] in combination with the linear complexity of SpeakEasy make it feasible to derive consensus clustering estimates, by clustering networks many times using differential initial conditions, to identify stable communities and multi-community nodes. Even on large full matrices, such as coexpression networks, the efficiency of SpeakEasy enables stable community estimates using typical hardware.

*Selecting final partitions and defining multi-community nodes:*
Stochastic clustering techniques such as SpeakEasy, have the potential to generate more accurate and robust results than typical methods that output a single set of communities [7,8]. Because SpeakEasy generates many "partitions" (sets of communities) during replicate runs, we are faced with the task of combining these outputs into a final partition.  This process is known as consensus clustering and in many cases solutions to this problem are more statistically challenging than basic clustering, essentially because it involves relationships of semi-overlapping sets.  The manner in which the "best" final partition is chosen among many stochastically generated partitions is independent of the method used to generate the partitions.  Therefore, improving consensus clustering could lead to better SpeakEasy results.

Overlapping communities and multi-community nodes are defined through a combination of the final representative partition, and the co-occurrence matrix, which is generated by all partitions.  The final representative partition is chosen to be the one with the highest average ARI with all other partitions.  Individual nodes with significant co-occurrence weight in more than one community can be classified as multi-community nodes.  Each entry A(i,j) of the co-occurrence matrix denotes how many times nodes *i* and *j* cluster together, under replicate SpeakEasy partitions.  We set the threshold for community membership as a function of the maximum number of community memberships considered realistic in a particular biological setting, specifically 1/max-number-of-communities.  This adaptive threshold is intuitive, because as the number of communities associated with a given node increases, the frequency in which a node is found in any one of the communities decreases.  This threshold is used in conjunction with the co-occurrence matrix: nodes that co-occur in multiple communities with an average weight (across all members of that community) greater than the threshold are identified as multi-community nodes.  More or less stringency in this threshold provides control over the number of multi-community nodes.  This flexibility in defining multi-community nodes is useful, because in some experimental settings it might be desirable to obtain more or less conservative definitions of multi-community nodes.

**Application-specific methods and data acquisition**

*Protein-protein interaction network processing*
Protein complexes were downloaded from various databases such as MIPS and these gold-standard complexes were compared to communities derived from binary interactions identified in the Gavin et al. [9] and Collins et al. [10].  Because we validate the clustering output by comparing to ground truth complexes, we operate on the (large) subsets of Gavin and Collins networks wherein the nodes on both sides of a given link are present in the ground truth.  If we did not do this, there would be nodes in the network for which we do not have ground truth community information.

*Application to sorted cell type populations*
We compare SpeakEasy results to several hierarchical clustering techniques that are often "first pass" clustering methods applied to cell type datasets such as Immgen.  While hierarchical techniques generate multiple levels of communities, it is unclear where to "cut" the hierarchical "tree" of communities in order to extract optimum communities and/or subcommunities.  Because the optimal number of communities in a dataset is generally unknown, we estimate the number of cell types in Immgen, using ten different methods [11]. When applied to Immgen, most of these methods proposed a single community containing all cells, or placed every cell in its own community. Therefore, we used the mode of estimates that do not fall at these extremes (recommendation: two communities). When using the recommended number of communities, all forms of hierarchical clustering showed relatively low correspondence with known cell types. We also test results when hierarchical methods are supplied with the true number of communities (information that is rarely available). Also, we compare results from single, average and complete hierarchical linkage methods. When compared to known cell classes (B cell, natural killer cells etc) SpeakEasy communities show higher NMI with the standard Immgen definitions than any hierarchical method, even when those methods are supplied with the true number of communities. The second iterative application of SpeakEasy also shows higher NMI with the nested classification of cell class with tissue of origin. In all cases it is much more accurate than when using the predicted number of true communities, which is a more realistic scenario. Moreover, results from single versus average or complete linkage vary substantially in their ability to recover true communities, but for a typical dataset where the true communities are unknown, the best linkage method is rarely known in advance.

*Gene coexpression network generation and assessment*
Because communities containing thousands of genes are difficult to test experimentally, we apply SpeakEasy iteratively three times to these datasets to reduce the typical community size to a few hundred genes, which is experimentally tractable. This might seem at odds with SpeakEasy's ability to automatically select community number. However, unlike hierarchical clustering, SpeakEasy cannot be forced to output subcommunities, because it is always admissible for the algorithm to place all nodes into a single community. To calculate median Bonferroni-corrected p-values for functional enrichment, we consider the top-scoring category for each module in a gene ontology biological process, among all modules with at least 30 genes (which is a practical threshold for GO enrichment) with at least one category scoring an un-corrected enrichment score of $p<.01$.

The weighted gene networks coexpression network analysis (WGNCA) tool is frequently used to identify coexpressed gene sets [12,13]. There are a significant number of practical and performance measures that distinguish SpeakEasy from this older method. For instance, generating communities with WGCNA entails (i) fitting a scale-free distribution, which may not be appropriate for all datasets, (ii) filtering link weights, and then (iii) performing a hierarchical treecut to select communities, a process which is notoriously sensitive to noise [14]. Indeed a supplementary routine for WGCNA with 19 parameters is offered to manually "tune" the hierarchical treecut for good results [15]. On the other hand, for this application to coexpression networks, SpeakEasy does not require any link filtering or distribution fitting and uses just a single parameter (number of times to sub-cluster). Furthermore, SpeakEasy gene sets are backed by the best recorded performance on the LFR benchmarks. In comparison, WGNCA cannot even be tested on the standard LFR benchmarks because it has so many manual tuning requirements. Functional enrichment scores for the top ten most enriched gene sets are either identical or highly similar for both methods, in both HBA and CCLE datasets, with larger differences in functional enrichment for smaller modules, with overall lower enrichment scores. The median Bonferroni-corrected p-values for functional enrichment across all modules are of equal magnitude on the HBA dataset, while they are 3-9 orders of magnitude more enriched for SpeakEasy communities derived from CCLE.

*Electrophysiology methods*
To simulate multichannel neural recordings from which to extract action-potentials (spikes) to test spike sorting, previously sorted spike waveforms [16] were added to brain noise (300-7kHz bandpass, 6.92uV RMS) to create a simulated 54 channel time series from a depth electrode array (spanning ~1mm with sites spaced every ~50um). 43 unique spike samples were added to the noise file 225 times (i.e. generating 43 communities of 225 elements, as well as one pseudo-community of 225 elements composed purely of noise). The waveform around each spike consists of a 1ms window (25 sampled time points) centered around the peak amplitude. A null class was included to simulate the presence of spurious spike detections that may occur when extracting spike waveform samples. We find that while the primary clusters detected by SpeakEasy provide good results, they can be improved further by applying it to each cluster, to detect additional subclusters (Table S1). Additional iterative subclustering does not improve results, indicating that only one round of subclustering is necessary for maximum accuracy.

*fMRI processing*
This analysis includes 27 subjects with PD ($M_{age}$= 66.52) and 21 controls ($M_{age}$ = 62), selected from a larger study [17]. Potential participants were excluded if they had a history of any primary neurodegenerative disease other than idiopathic Parkinson Disease, a history of brain surgery for PD, moderate to severe dyskinesia, significant head trauma, stroke history, severe or unstable cardiovascular disease, contraindications to MRI, or a Montreal Cognitive Assessment score (MoCA) [18] lower than 23. This study was approved by the University of Washington Institutional Review Board. All participants provided written informed consent.

See Table S1 for sample characteristics. Participants were predominantly right-handed. PD patients did not differ significantly from controls on age, education, or scores on the MoCA. The PD subjects have declined from their premorbid cognitive abilities, but they are well within the range of normative values for normal controls [18]. PD subjects had significantly higher scores on the UPDRS motor subscale [19].

Males were over-represented in the PD group, consistent with higher incidence rates of Parkinson disease in men [20]. PD patients ranged from Hoehn and Yahr stage 1 to 2.5 with most at stage 2 (N=17). At the time of the scan and corresponding neuropsychological evaluations, most PD patients were taking dopaminergic medications (24% were taking both levodopa and a dopamine agonist, 38% were taking only levodopa, 19% were taking only a dopamine agonist, and 19% were taking no dopaminergic medications).

*MRI acquisition*
Scans were performed after morning doses of dopaminergic medication (if applicable). Data were acquired using a Philips 3T Achieva MR System (Philips Medical

Systems, Best, Netherlands, software version R2.6.3) with a 32-channel SENSE head coil.  During each session, whole-brain axial echo-planar images (43 sequential ascending slices, 3 millimeter isotropic voxels, field of view = 240x240x129 , repetition time = 2400 ms, echo time = 25 ms, flip angle = 79°, SENSE acceleration factor = 2) were collected parallel to the AC-PC line for a single resting state run and six task runs. Run duration was 300 volumes (12 minutes) for the resting state run which we split into two volumes of 6 minutes from each patient.  The rationale for this was to enhance our ability to avoid temporally smearing among evolving brain region communities while remaining within the resting-state paradigm.  A sagittal T1-weighted 3D MPRAGE (176 slices, matrix size = 256 x 256, inversion time = 1100 ms, turbo-field echo factor = 225, repetition time = 7.46 ms, echo time = 3.49 ms, flip angle =7°, shot interval = 2530 ms) with 1 mm isotropic voxels was also acquired for registration.

*MRI processing*

Functional images from rest or task were processed identically using a pipeline developed using software from FSL [21], FreeSurfer [22], and AFNI [23]. Data were corrected for motion using FSL MCFLIRT (M. Jenkinson, Bannister, Brady, & Smith, 2002). The pipeline removed spikes using AFNI, performed slice timing correction using FSL, and regressed out time series motion parameters and the mean signal for eroded (1mm in 3D) masks of the lateral ventricles and white matter (derived from running FreeSurfer on the T1-weighted image). We did not regress out the global signal. We did not perform bandpass filtering to avoid artificially inflating correlations or inducing structure that was not actually present in the data, and because resting state networks exhibit different levels of phase synchrony at different frequencies (Handwerker, Roopchansingh, Gonzalez-Castillo, & Bandettini, 2012; Niazy, Xie, Miller, Beckmann, & Smith, 2011). Three dimensional spatial smoothing was performed using a Gaussian kernel with a FWHM of sigma=3mm. Co-registration to the T1 image was performed using boundary based registration based on a white matter segmentation of the T1 image (epi_reg in FSL).

We selected 264 MNI coordinates from a previous partitioning of fMRI data into functional nodes by Power et al.  [24]. For each coordinate, we created a 10mm diameter mask in standard space and transformed that to subjects' native space to calculate mean subject-specific timecourses for each ROI. We calculated the Pearson correlation between each pair of nodes to obtain an edge weight, representing the strength of connectivity.

*Significance of connectivity changes*

Significance of changes in community membership between resting state networks from the control cohort compared to the Parkinson's cohort were estimated through permutation tests.  We (repeatedly) randomly mix control and Parkinson's samples to generate two average connectivity matrices as pseudo-disease and pseudo-control groups.  There should be no disease-related differences between these matrices generated from randomized disease and control data.  We generate such mixed disease+control connectivity matrices 1000 times and detect communities in each case, generating 2000 paired (pseudo-disease/pseudo-control) co-occurrence matrices.  The distribution of differences between these co-occurrence matrices are used to generate a null distribution used to estimate the significance of changes within and between communities between the real control and PD co-occurrence matrices.

We compare these clustering results to those generated by Infomap [25], a clustering algorithm based on data compression that has previously been used in fMRI analysis [24]. Infomap cannot take advantage of negative links found in correlations matrices, so we follow the standard practice of converting all links to positive values. We find that Infomap places all brain regions into a single community, unless links are extensively filtered. After correlations with R-value < 0.75 are removed, communities begin to emerge. However, these communities are not robust; for instance, the difference between various Infomap-based partitions (at different correlation thresholds) is as large as the difference between the control and PD partitions, using SpeakEasy (Table S2). Furthermore, Infomap communities are a point estimate, and there is no measure of the robustness of communities at a particular threshold.

**Figure S1 Robust clustering performance with various community size distributions and intra-community degree distributions.** (A) Various disjoint community recovery metrics for networks from LFR benchmarks with n=1000, $\gamma$ (community size distribution) =3, $\beta$ (within-community degree distribution) =2. (B) Disjoint community recovery metrics for networks from LFR benchmarks with n=1000, $\gamma$=3, $\beta$=1 (C) Disjoint community recovery metrics for networks from LFR benchmarks with n=1000, $\gamma$=2, $\beta$=2.

**Figure S2 Brain region communities detected from control subject resting-state fMRI.** The order of communities 1-6 corresponds to the order of communities shown in Figure 5. Location of brain regions in each cluster was/were visualized with the BrainNet Viewer [26].

**Table S1. Comparison of communities of similar neuronal spikes vs known spike communities.**

**Table S2. Comparison of brain region communities detected in control or PD cohorts using SpeakEasy or Infomap (the later using various thresholds for link significance).**

**Table S3. Summary of demographics of control and PD cohorts in resting-state fMRI study.**

Figure S1

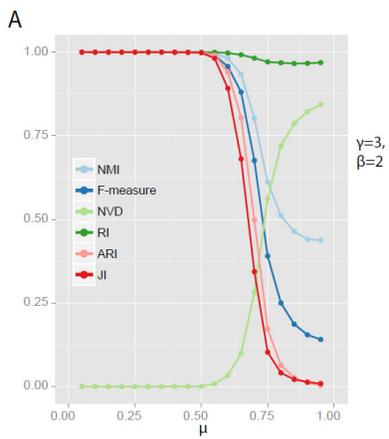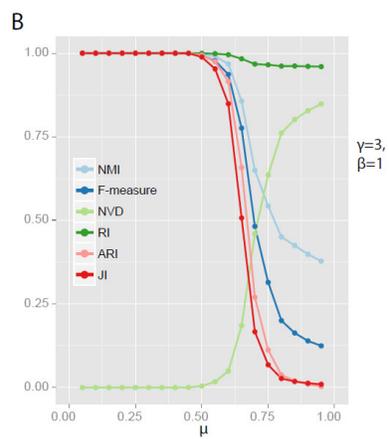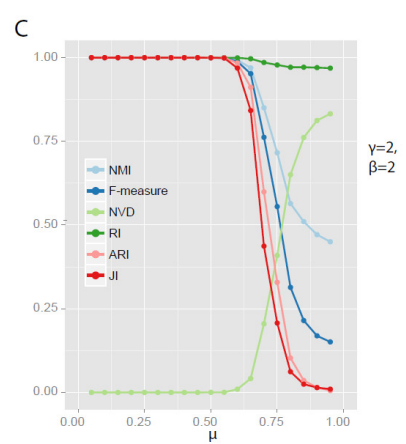

Figure S2

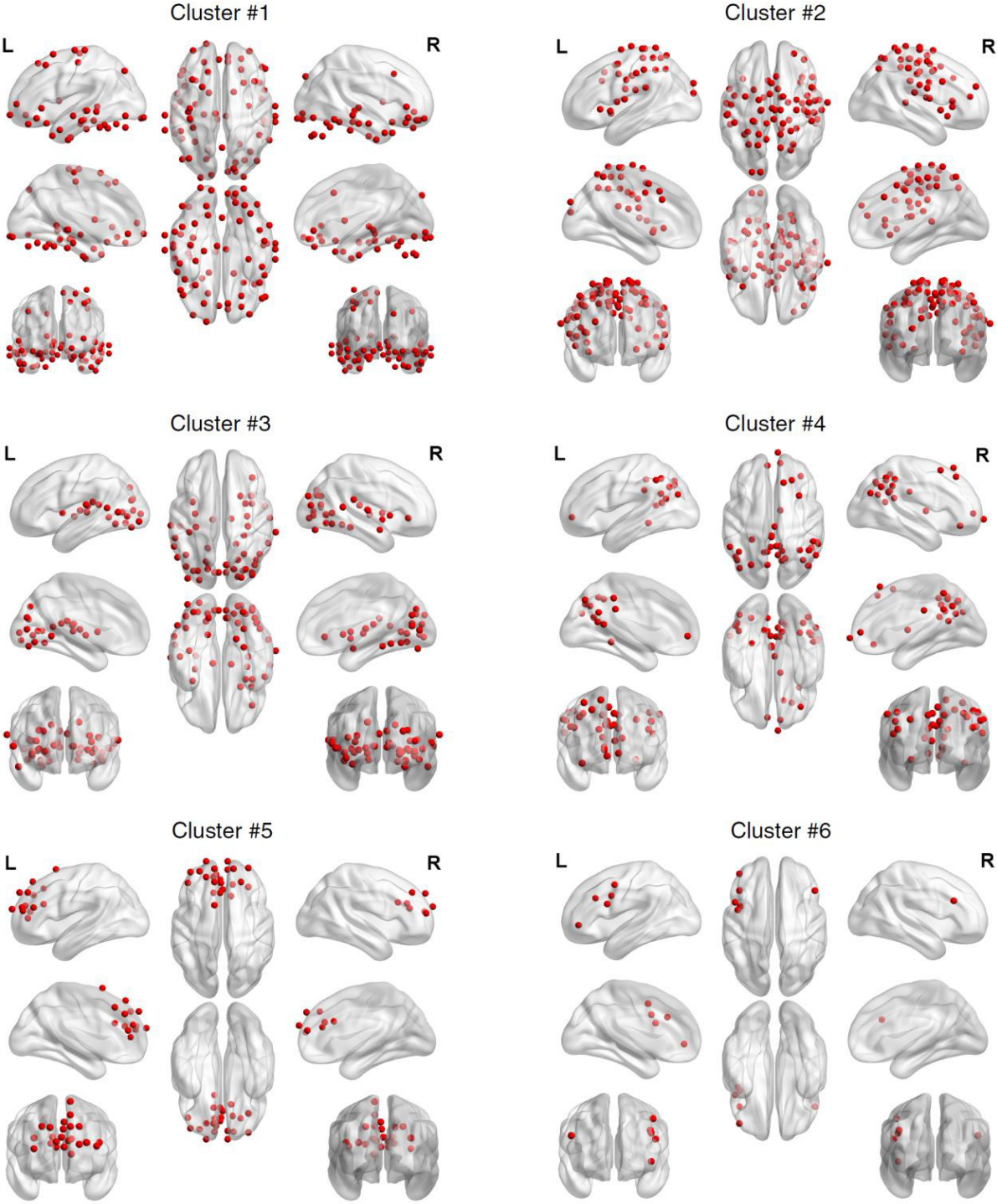

## Table S1

| NMI | SpeakEasy primary communities | SpeakEasy secondary communities | SpeakEasy tertiary communities |
|---|---|---|---|
| spike waveform dataset 1 | 0.6381 | 0.8328 | 0.8085 |
| spike waveform dataset 2 | 0.6250 | 0.8202 | 0.7793 |
| spike waveform dataset 3 | 0.6127 | 0.8178 | 0.8052 |
| mean | 0.6253 | 0.8236 | 0.7977 |

| adjusted Rand index | SpeakEasy primary communities | SpeakEasy secondary communities | SpeakEasy tertiary communities |
|---|---|---|---|
| spike waveform dataset 1 | 0.6455 | 0.8274 | 0.8103 |
| spike waveform dataset 2 | 0.5981 | 0.7922 | 0.7772 |
| spike waveform dataset 3 | 0.6487 | 0.8251 | 0.8066 |
| mean | 0.6308 | 0.8149 | 0.7980 |

## Table S2

|  | PD | Control | Total |
|---|---|---|---|
| N | 27 | 21 | 48 |
| Age at Scan | 66.52( 9.86) | 61.90(10.00) | 64.50(10.08) |
| Sex(number males) | 20(74%) | 9 (43%) | 29 (60%) |
| Education (years) | 16.35(2.10) | 15.90(2.39) | 16.15(2.22) |
| Hoen & Yahr | 2.03 (1–2.5) |  |  |
| Handedness (Right) | 23 | 19 | 42 |
| Dominant side of motor symptoms | 7 Left/ 18 Right/ 2 Symmetric |  |  |
| UPDRS Part I | 10.00(5.51) |  | 10.00(5.51) |
| UPDRS Part II | 8.81(5.35) |  | 8.81(5.35) |
| UPDRS Part III | 23.30( 8.49) | 0.81( 1.40) | 13.46(12.95) |
| UPDRS Part IV | 1.85(3.59) |  | 1.85(3.59) |
| Levodopa (current) | 18 | 0 | 18 |
| Dopamine agonist (current) | 11 | 0 | 11 |
| Years since symptom onset | 8.71(5.01) |  |  |
| MOCA | 26.44(2.06) | 27.29(1.95) | 26.81(2.04) |
| Hopkins Verbal Learning Test | 24.48(5.69) |  |  |
| Golden Stroop (total correct) | 189.26(24.99) |  |  |
| Trails B (seconds) | 74.42(31.53) |  |  |

## Table S3

| NMI between partitions | Control SpeakEasy | PD SpeakEasy | Control Infomap >0.75 | Control Infomap >0.80 | Control Infomap >0.85 | Control Infomap >0.90 | Control Infomap >0.95 | PD Infomap >0.75 | PD Infomap >0.80 | PD Infomap >0.85 | PD Infomap >0.90 | PD Infomap >0.95 |
|---|---|---|---|---|---|---|---|---|---|---|---|---|
| Control SpeakEasy | 1 | 0.5141 | 0.1218 | 0.3285 | 0.5376 | 0.5952 | 0.604 | 0.0866 | 0.2319 | 0.4426 | 0.4863 | 0.5057 |
| PD SpeakEasy | 0.5141 | 1 | 0.0701 | 0.3552 | 0.4569 | 0.5897 | 0.6491 | 0.0609 | 0.3165 | 0.6957 | 0.6919 | 0.633 |
| Control Infomap >0.75 | 0.1218 | 0.0701 | 1 | 0.2122 | 0.144 | 0.1133 | 0.0989 | 0.4727 | 0.1506 | 0.094 | 0.0912 | 0.0974 |
| Control Infomap >0.80 | 0.3285 | 0.3552 | 0.2122 | 1 | 0.6043 | 0.3834 | 0.3542 | 0.108 | 0.5029 | 0.3618 | 0.3586 | 0.3266 |
| Control Infomap >0.85 | 0.5376 | 0.4569 | 0.144 | 0.6043 | 1 | 0.6502 | 0.5291 | 0.0829 | 0.3364 | 0.5114 | 0.4686 | 0.473 |
| Control Infomap >0.90 | 0.5952 | 0.5897 | 0.1133 | 0.3834 | 0.6502 | 1 | 0.702 | 0.1368 | 0.2527 | 0.6387 | 0.6605 | 0.6829 |
| Control Infomap >0.95 | 0.604 | 0.6491 | 0.0989 | 0.3542 | 0.5291 | 0.702 | 1 | 0.1225 | 0.3306 | 0.6676 | 0.7141 | 0.808 |
| PD Infomap >0.75 | 0.0866 | 0.0609 | 0.4727 | 0.108 | 0.0829 | 0.1368 | 0.1225 | 1 | 0.3245 | 0.1988 | 0.1929 | 0.1626 |
| PD Infomap >0.80 | 0.2319 | 0.3165 | 0.1506 | 0.5029 | 0.3364 | 0.2527 | 0.3306 | 0.3245 | 1 | 0.4899 | 0.4753 | 0.371 |
| PD Infomap >0.85 | 0.4426 | 0.6957 | 0.094 | 0.3618 | 0.5114 | 0.6387 | 0.6676 | 0.1988 | 0.4899 | 1 | 0.8803 | 0.7164 |
| PD Infomap >0.90 | 0.4863 | 0.6919 | 0.0912 | 0.3586 | 0.4686 | 0.6605 | 0.7141 | 0.1929 | 0.4753 | 0.8803 | 1 | 0.7559 |
| PD Infomap >0.95 | 0.5057 | 0.633 | 0.0974 | 0.3266 | 0.473 | 0.6829 | 0.808 | 0.1626 | 0.371 | 0.7164 | 0.7559 | 1 |